\documentclass[preprint,12pt]{elsarticle}

\usepackage{amssymb}

\usepackage{graphicx}
\usepackage{dcolumn}
\usepackage{bm}
\usepackage{epsfig}

\journal{Physics Letters B}

\begin{document}
\def\d{{\rm d}}
\def\Epos{E_{\rm pos}}
\def\ap{\approx}
\def\eff{{\rm eft}}
\def\L{{\cal L}}
\newcommand{\vev}[1]{\langle {#1}\rangle}
\newcommand{\CL}   {C.L.}
\newcommand{\dof}  {d.o.f.}
\newcommand{\eVq}  {\text{EA}^2}
\newcommand{\Sol}  {\textsc{sol}}
\newcommand{\SlKm} {\textsc{sol+kam}}
\newcommand{\Atm}  {\textsc{atm}}
\newcommand{\Chooz}{\textsc{chooz}}
\newcommand{\Dms}  {\Delta m^2_\Sol}
\newcommand{\Dma}  {\Delta m^2_\Atm}
\newcommand{\Dcq}  {\Delta\chi^2}
\newcommand{\nbb}{$\beta\beta_{0\nu}$ }
\def\VEV#1{\left\langle #1\right\rangle}
\let\vev\VEV
\def\e6{E(6)}
\def\10{SO(10)}
\def\21{SA(2) $\otimes$ U(1) }
\def\321{$\mathrm{SU(3) \otimes SU(2) \otimes U(1)}$ }
\def\lr{SA(2)$_L \otimes$ SA(2)$_R \otimes$ U(1)}
\def\422{SA(4) $\otimes$ SA(2) $\otimes$ SA(2)}
\newcommand{\AHEP}{%
School of physics, Institute for Research in Fundamental Sciences
(IPM)\\P.O.Box 19395-5531, Tehran, Iran\\
  Department of Physics, Sharif University of Technology\\
  P.O.Box 11155-9161, Tehran, Iran
  }
\newcommand{\Tehran}{%
School of physics, Institute for Research in Fundamental Sciences
(IPM)
\\
P.O.Box 19395-5531, Tehran, Iran}
\def\roughly#1{\mathrel{\raise.3ex\hbox{$#1$\kern-.75em
      \lower1ex\hbox{$\sim$}}}} \def\lsim{\roughly<}
\def\gsim{\roughly>}
\def\ltap{\raisebox{-.4ex}{\rlap{$\sim$}} \raisebox{.4ex}{$<$}}
\def\gtap{\raisebox{-.4ex}{\rlap{$\sim$}} \raisebox{.4ex}{$>$}}
\def\lsim{\raise0.3ex\hbox{$\;<$\kern-0.75em\raise-1.1ex\hbox{$\sim\;$}}}
\def\gsim{\raise0.3ex\hbox{$\;>$\kern-0.75em\raise-1.1ex\hbox{$\sim\;$}}}
\begin{frontmatter}


\author{Y. Farzan\corref{cor1}}
\ead{yasaman@theory.ipm.ac.ir}
 \address{School of physics, Institute for Research in
Fundamental Sciences (IPM), P.O.Box 19395-5531, Tehran, Iran}
\author{A. Rezaei Akbarieh\corref{cor2}}
\ead{am$_-$rezaei@physics.sharif.ir}
\address{School of physics, Institute for Research in Fundamental Sciences
(IPM),P.O.Box 19395-5531, Tehran, Iran}
\address{Department of Physics, Sharif University of Technology, P.O.Box 11155-9161, Tehran, Iran}
\title{Natural explanation for 130 GeV photon line within vector boson dark matter
model}




\begin{abstract}
We present a dark matter model for explaining the observed 130~GeV
photon line from the galaxy center. The dark matter candidate is a
vector boson of mass $m_V$ with a dimensionless coupling to the
photon and $Z$ boson. The model predicts a double line photon
spectrum
 at energies equal to
$m_V$ and $m_V(1-m_Z^2/4m_V^2)$ originating from the dark matter
annihilation. The same coupling leads to a mono-photon plus missing
energy signal at the LHC. The entire perturbative parameter space
can be probed by the 14 TeV LHC run. The model has also a good
prospect of being probed by direct dark matter searches as well as
the measurement of the rates of $h \to \gamma \gamma$ and $h \to Z
\gamma$ at the LHC.
\end{abstract}


\end{frontmatter}


\section*{Introduction}
Recently a monochromatic photon line at 130 GeV has been found in
 the FermiLAT data in the vicinity of the galactic center
 \cite{130line,gamma+Z}.  A possible explanation for this line can
 be the annihilation of a Dark Matter (DM) pair of mass 130 GeV directly
 to a photon pair with cross section equal to $10^{-37}~{\rm
 cm}^2$. Extensive studies have been carried out in the literature
 to explain this line \cite{mambrini,extensive}. In these models, the DM is
 taken to be either a scalar or a fermion so the annihilation to a
 photon pair cannot  take place at a tree level   with renormalizable couplings. If the charged particles propagating
 in the loop are light enough, their direct production via DM annihilation
  typically exceeds the bounds \cite{three-exceptions}.

   However, within Vector Dark Matter (VDM) models novel features appear.
     Such VDM models have recently
    received attention in the literature \cite{vector,us}. Here, we
  show that the vector boson DM candidate has a unique advantage
  for explaining the 130 photon line because unlike a neutral scalar or spinor, a neutral
   vector boson can directly
  couple to photon through a large unsuppressed dimensionless gauge invariant
  coupling. In this letter, we introduce a simple model that explains the 130 GeV
  line. The model predicts accessible new signals for the LHC and
  can explain the slight excess of $Br(h\to \gamma \gamma)$. In
  sec. \ref{model}, we introduce the model and discuss the direct and
  indirect DM searches within this model. In sec.
  \ref{excess}, we compute the contribution to the Higgs decay to
  a  photon pair. In sec. \ref{LHC}, we discuss the potential
  signal at colliders. Results are summarized in sec. \ref{con}.
\section{The model \label{model}}
\begin{figure}[t]
\vspace{-9mm} \hspace{30mm}
\includegraphics[width=70mm]{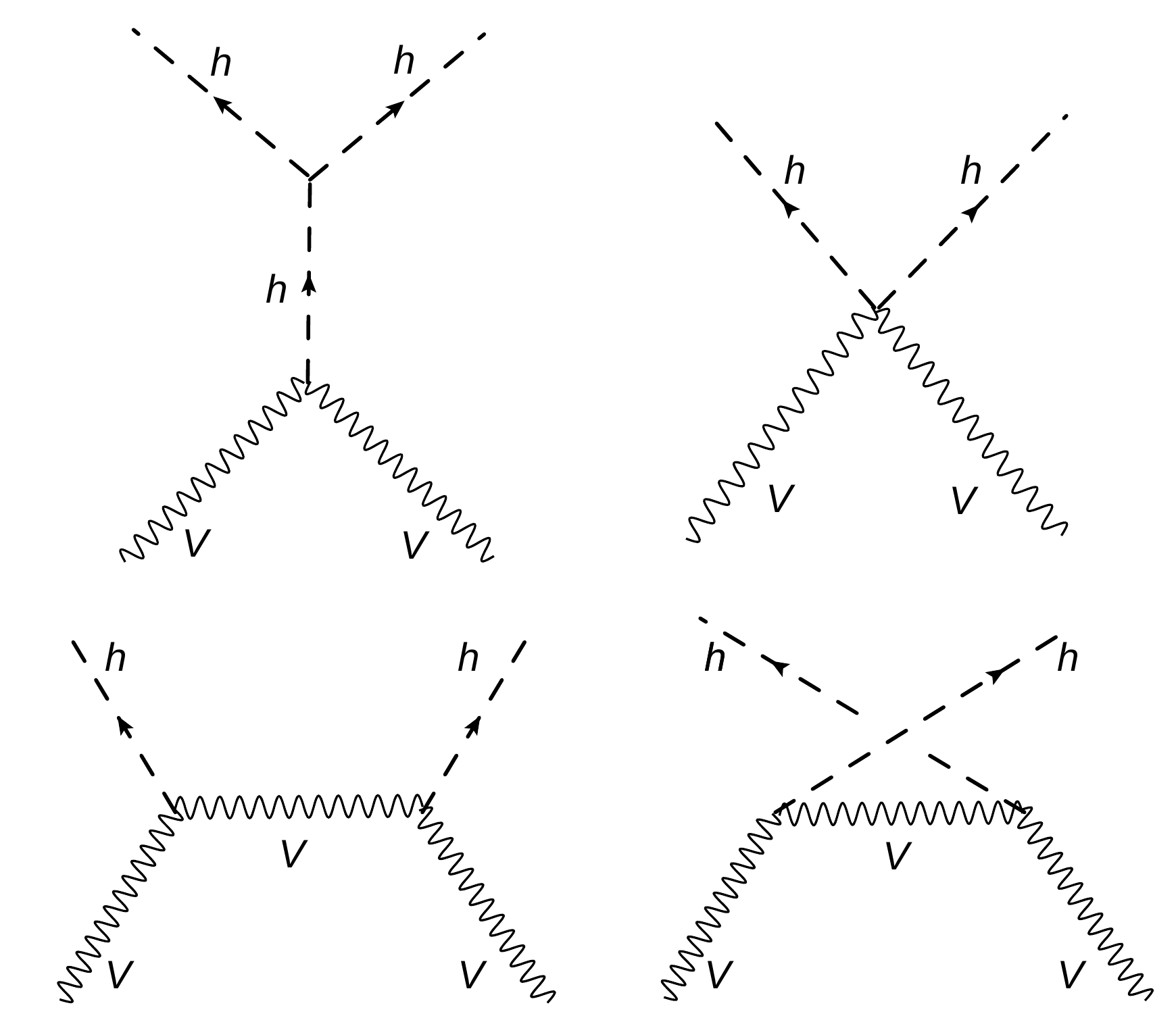}\caption{Annihilation  of the $V$ pair
to a Higgs pair via $\lambda_1$ coupling }
\end{figure}\
The model adds only a pair of neutral vector bosons $V$ and
$V^\prime$ with masses $m_V<m_{V^\prime}$ to the Standard Model
(SM). We impose a $Z_2$ symmetry under which only $V$ and $V^\prime$
are odd so $V$ is stable and therefore a potentially suitable dark
matter candidate. To avoid negative norm modes, we take their
kinetic terms to be of antisymmetric form: {\it i.e.,}
$-\left[V_{\mu \nu}V^{\mu \nu}+V_{\mu \nu}^\prime V^{\prime\mu
\nu}\right]/4$ where $V_{\mu \nu}^{(\prime)}\equiv
\partial_\mu V_\nu^{(\prime)} -\partial_\nu V_\mu^{(\prime)}$.
In a general basis, the mass terms are \begin{equation}
\frac{m_1^2}{2} V\cdot V +\frac{m_2^2}{2} V^\prime\cdot
V^{\prime}+m_3^2V\cdot V^\prime \end{equation} and the dimensionless
gauge and $Z_2$ invariant couplings to the Higgs are
\begin{equation} \label{HV} \frac{\lambda_1}{2}|H|^2 V_\mu V^\mu
+\frac{\lambda_2}{2}|H|^2 V_\mu^\prime V^{\prime\mu}+\lambda_3|H|^2
V_\mu^\prime V^{\mu}.\end{equation} Without loss of generality we
can go to  a basis in which $V$ and $V^\prime$ are mass eigenstates
with masses $m_V^2$ and $m_{V^\prime}^2$. Notice that in this basis
$\lambda_3$ can be nonzero but $\lambda_3 v_h^2/2+m_3^2=0$. Although
the $\lambda_i$ couplings are dimensionless, if $V_\mu$ are not
gauge bosons, they will be non-renormalizable \cite{Weinberg}. We
can promote $V_\mu$ and $V_\mu^\prime$ to gauge bosons of two new
$U(1)$ gauge symmetries as prescribed in the St\"uckelberg
mechanism, by replacing $V_\mu$ and $V_\mu^\prime$ in these terms as
well as in mass terms with $\partial_\mu \theta_V-V_\mu$ and
$\partial_\mu \theta_{V^\prime}-V_\mu^\prime$. We will work in a
gauge that the St\"uckelberg fields $\theta_V$ and
$\theta_{V^\prime}$ are eaten by the longitudinal components of $V$
and $V^\prime$. For the purpose of this paper, it is enough to take
$\lambda_i$ Wilsonian effective couplings below some cutoff
$\Lambda$.
 The new
vector boson can have quartic couplings with each other but these
couplings are not relevant for our discussion. Notice that
$\lambda_i$ should be real to guarantee the Hermiticity of the
potential; however, because of the presence of quartic vector boson
coupling, we do not know a priori their sign.

 The $\lambda_1$
and $\lambda_3$ couplings give rise to annihilation of to the $V$
pair (see Figs 1,2).
Setting $\lambda_3=0$, we find\\
\begin{eqnarray}
\langle\sigma(VV\to
hh)v_{rel}\rangle&=&\frac{\lambda_1^2\sqrt{m_V^2-m_h^2}}{576\pi
m_V^3}\nonumber
\end{eqnarray}
\begin{eqnarray}
\{\frac{[(4m_V^2-m_h^2)(m_V^2+2\lambda_1v_h^2)-\frac{3}{2}\tan\theta_Wm_V^2m_h^2]^2}{m_V^4(m_H^2-4m_V^2)^2}\nonumber\\
+2[1+\frac{2\lambda_1 v_h^2}{2m_V^2-m_h^2}+\frac{3\tan\theta_W
m_h^2}{2(-4m_V^2+m_h^2)}]^2\}
\end{eqnarray}
The vacuum expectation value of Higgs is denoted by $v_h=246$ GeV.
Moreover, like other Higgs portal models, the $\lambda_1$ coupling
gives rise to the annihilation of the $V$ pair via an $s$-channel
Higgs exchange diagram with cross section
 \begin{eqnarray}
\langle\sigma (VV\to f \bar{f})v_{rel}\rangle&=&
\frac{\lambda_1^2v_h^2\Gamma(h^*\to f \bar{f})}{3
m_V(4m_V^2-m_h^2)^2},\nonumber
\end{eqnarray}
where $f \bar{f}$ can be $W^+ W^-$, $ZZ$, $b \bar{b}$ and etc.
$\Gamma(h^*\to f \bar{f})$ is the decay rate of a hypothetical
SM-like Higgs ($h^*$) with a mass equal to $2m_V$ to $f \bar{f}$.
\begin{figure}[t]
\vspace{-9mm} \hspace{30mm}
\includegraphics[width=70mm]{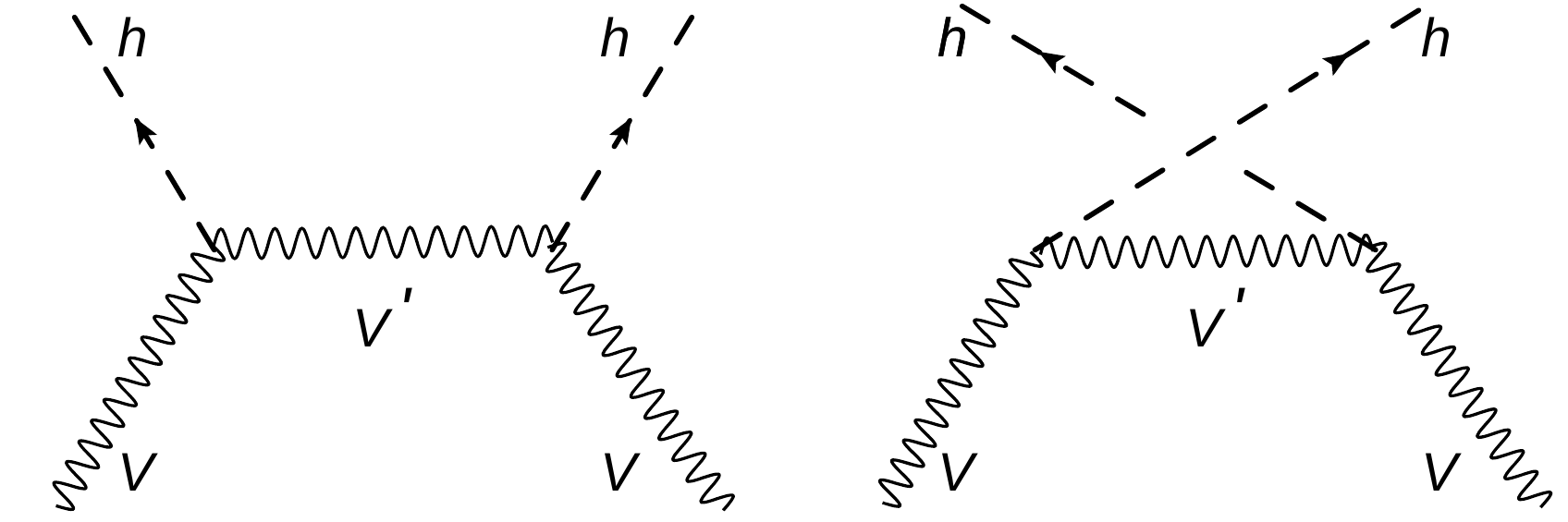}\caption{Annihilation  of the $V$ pair
to a Higgs pair via the $\lambda_3$ coupling}
\end{figure}\
To account for the observed dark matter abundance within the thermal
production scenario \cite{Gel}, the total DM pair annihilation cross
section should be equal to  1~pb. Setting the sum of cross sections
of these  modes equal to 1~pb and $m_V=130$~GeV, we find
$\lambda_1=0.09$. Notice that the total annihilation  cross section
falls well below the $10^{-25} cm^3/s$ bound from Fermi-LAT
continuum gamma-ray constraint \cite{Cohen} as well as the bounds
from the PAMELA constraint on the anti-proton flux \cite{Cirelli}.
More data from Fermi-LAT and AMS02 may make it possible to probe the
model in future.
 The $\lambda_3$ coupling  also gives rise to annihilation to a Higgs pair via a $t-$ and $u-$ channel
 $V^\prime$ exchange (see Fig.2). Fixing $\lambda_1=0$, we find
\begin{eqnarray}
\langle\sigma(VV\to
hh)v_{rel}\rangle=\frac{\lambda_3^4v_h^4\sqrt{m_V^2-m_h^2}}{144\pi
m_V^3}\nonumber\\
\{\frac{1}{m_{V^\prime}^4}+\frac{2}{(m_V^2-m_h^2+m_{V^\prime}^2)^2}\}.
\end{eqnarray}
Taking $\sigma(VV\to hh)=1$~pb,  for $\lambda_1=0$ and
$m_V=130$~GeV, we find $\lambda_3\simeq 0.4 (m_{V^\prime}/300~{\rm
GeV})$.
Notice that for  $\lambda_1=0$ and $m_{V^\prime}>O(3~{\rm TeV})$ the
required $\lambda_3$ enters the non-perturbative regime.
\begin{figure}[t]
\vspace{7mm} \hspace{30mm}
\includegraphics[width=70mm]{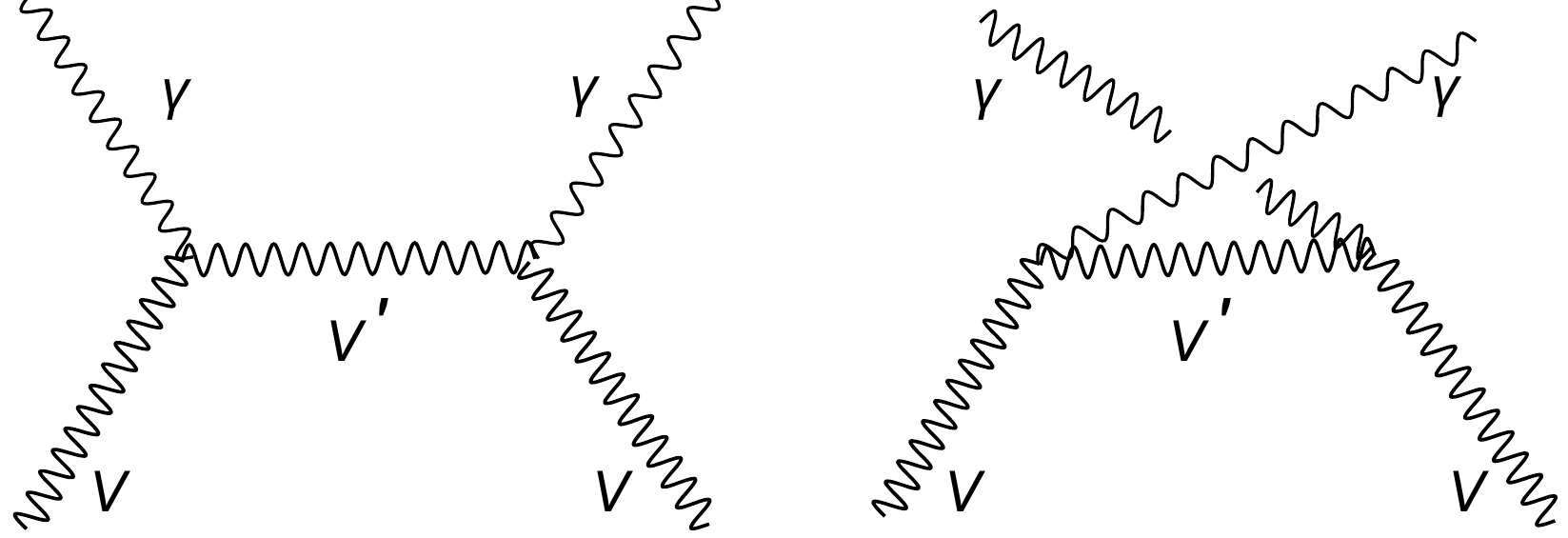}\caption{Annihilation of the $V$ pair
to a photon pair}
\end{figure}\
The symmetries of the model also allow the presence of the
following terms\footnote{ When revising the present paper, Ref.
\cite{over} appeared which has some overlap with our work.}:
\begin{equation} \label{gvs} g_V B^{\mu \nu} V_\mu V_\nu^\prime
+g^{\prime}_{V}\epsilon^{\mu\nu\alpha\beta}B_{\mu\nu}V_{\alpha}V_\beta^\prime
,\end{equation} where $B^{\mu \nu}$ is the field strength
associated with the hypercharge gauge  boson: $B_{\mu \nu}= \cos
\theta_W F_{\mu \nu}- \sin \theta_W Z_{\mu \nu}$. Again although
$g_V$ and \textbf{$g_V^\prime$} {are} dimensionless, if $V_\mu$
and $V_\mu^\prime$ are not promoted to gauge bosons, {these
couplings will be} non-renormalizable \cite{non-decoupling}. Again
using the St\"uckelberg mechanism, {these terms} can be made gauge
invariant. The $g_V$ coupling is the familiar ``generalized
Chern-Simons'' term \cite{Anastasopoulos:2006cz} which can arise
by integrating out heavy chiral fermions charged both under the
hypercharge and the new $U(1)$ gauge symmetries. The
\textbf{$g_V^\prime$} coupling can be large contrary to the
non-decoupling theorem \cite{non-decoupling}. A similar term has
also been employed in \cite{mambrini} to explain the 130 GeV line.
In the following, we study the phenomenology of these two terms.

{The $g_V$ and $g_V^\prime$ couplings} respectively lead to (see
Fig.~3)
\begin{eqnarray}
\langle\sigma(V+V\to \gamma
\gamma)v_{rel}\rangle&=&\frac{g_V^{(')4}\cos^4\theta_W}{9\pi}\frac{
R(x)}{m_V^2x^2(1+x)^2}\nonumber
\end{eqnarray} in which
$x=(m_{V^\prime}/m_V)^2$. For  the contribution from the $g_V$
coupling, $R(x)=(2+8x+9x^2)/8$  and  for that
 from the $g_V^\prime$ coupling, $R(x)=2(2+x^2)$. These couplings
also induce annihilation to a $Z\gamma$ pair as follows
$$\langle\sigma(V+V\to \gamma Z)v_{rel}\rangle=\frac{g_V^{(\prime)4}\sin^22\theta_W(4y-1)^3f^{(\prime)}(y,y^{'})}{9\pi
2^{12}m_Z^2y^{4}y^{'2}(1-2y-2y^{'})^2}$$\\
where
$$
f(y,y^\prime)=32y^4+y^{\prime 2}+16y^3(8y^{\prime}-3)
+6y^2(24y^{\prime 2}-16y^{\prime }+3)+y(8y^{\prime 2}-8y^{\prime
}+1)$$\\and\\$$ f^\prime(y,y^{'})=1/2+64y^4+32y^{\prime
2}-64y^3(1+16y^{\prime})+4y(64y^{\prime
2}-1)+16y^2(1+16y^{\prime}+160y^{\prime 2}),$$  in which
$y=(m_V/m_Z)^2$ and $y^\prime=(m_{V^\prime}/m_Z)^2$. We therefore
expect two photon lines: one photon line at $m_V$ and another at
$m_V(1-m_Z^2/4m_V^2)$ with an intensity suppressed by
$\sigma(V+V\to \gamma Z)/[2\sigma(V+V\to \gamma
\gamma)]<(\tan^2\theta_W)=0.3$  irrespective of the ratio
$g_V/g_V^\prime$. In fact, the observation favors double line
structure over a single line \cite{gamma+Z}; however, more data is
required to resolve such a double line feature \cite{laha}. From
now on, we take $m_V=130$ GeV and $\sigma(V+V\to \gamma
\gamma)=10^{-37}~{\rm cm}^2$.
For $m_{V^\prime}\geq 300$ GeV, this yields $g_V\simeq 0.27
(m_{V'}/300~{\rm GeV})$  for the $g_V$-dominated range and
$g_V^\prime\simeq 0.24 (m_{V'}/300~{\rm GeV})$   for the
$g_V^\prime$-dominated range. As long as $m_{V^\prime}<$a few TeV,
the required values of $g_V$ and $g_V^\prime$ will remain in the
perturbative regime.

Through the $\lambda_1$ coupling, the dark matter interacts with
nuclei with cross section  of $$\sigma_{SI}(V+N\rightarrow
V+N)=\frac{\lambda_1^2f^2}{4\pi}\frac{m_N^2m_r^2}{m_V^2m_h^4}$$\\
where $m_N$ is the mass of a nucleon and $m_r=m_Nm_V/(m_V+m_N)$ is
the reduced mass for the collision. $f$ parameterizes the nuclear
matrix element  ($0.14<f<0.66$) \cite{f}. Taking
$\lambda_1={0.09}$, we find $\sigma=4.4\times
10^{-45}(f/{0.27})^2{\rm cm}^2$. This means under the condition
that $\lambda_1$ is the main contributor to the DM annihilation,
the present bound from XENON100 \cite{xenon2012} practically rules
out $f>{0.27}$. However for $\lambda_1\ll\lambda_3\simeq
0.5~(m_{V'}/300~{\rm GeV})$,
we do not expect an observable effect in the direct searches. \\
 If the mass splitting between $V$ and $V^\prime$ is smaller than
 $O(100 ~{\rm keV})$, the DM can interact inelastically with the
 $g_V$ and {$g_V^\prime$ couplings} through a $t$-channel photon exchange. Small
 splitting can be justified by an approximate $ V\leftrightarrow
 V^\prime$ symmetry. We do not however consider such a limit so the main
 interaction will be via the Higgs portal channel.

\section{ Higgs decay to a photon pair \label{excess}}
\begin{figure}[t]
\vspace{-9mm} \hspace{30mm}
\includegraphics[width=70mm]{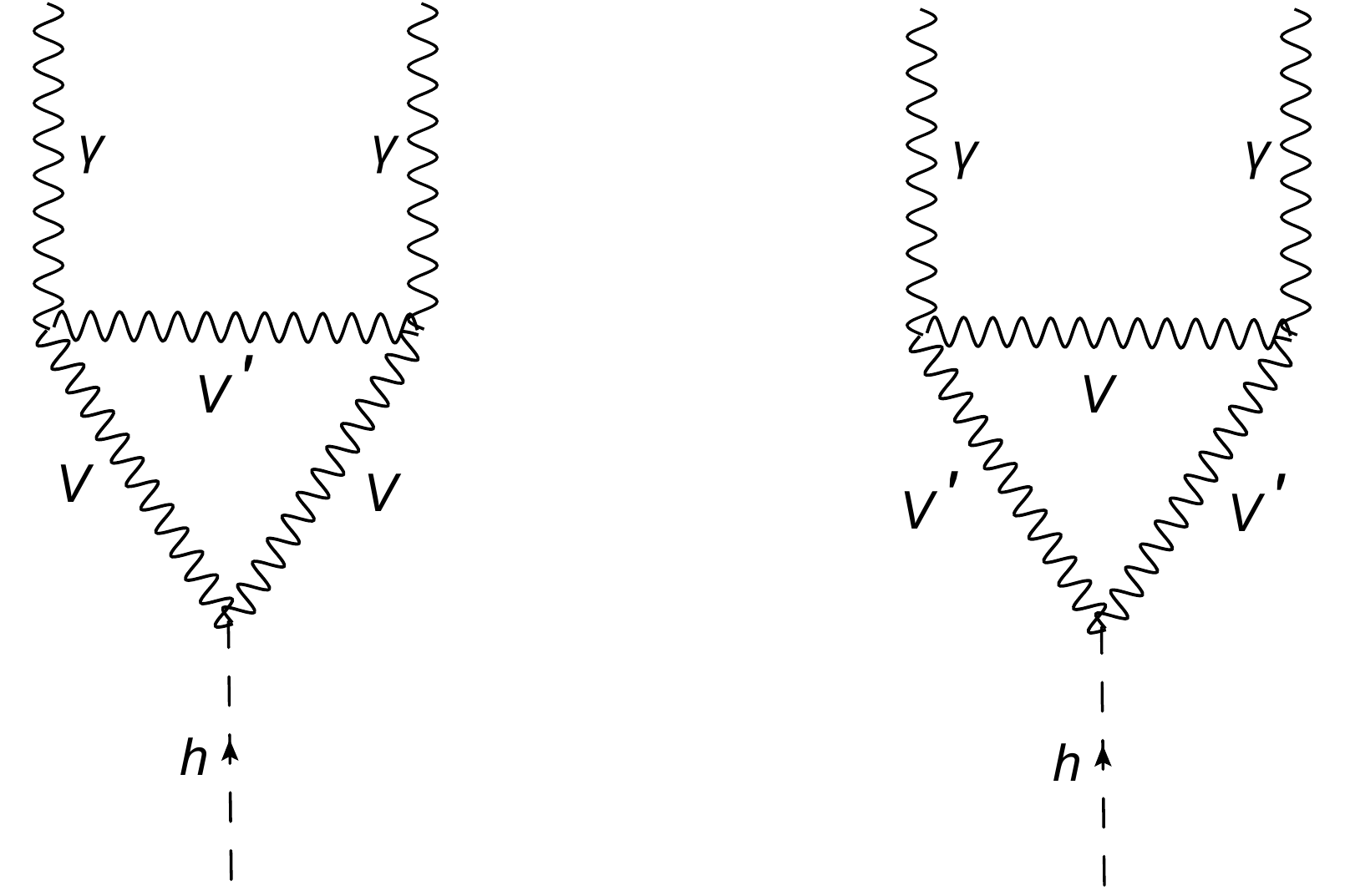}\caption{Diagrams of Higgs decay to a photon pair
via the $\lambda_1$ and $\lambda_2$ couplings}
\end{figure}\
 The $\lambda_1$ and  $\lambda_2$ couplings contribute to $h \to \gamma \gamma$
 via triangle
 diagrams within which $V$ and $V^\prime$ propagate (see Fig.4).
 For the $g_V$ and $g_V^\prime$ couplings,
 the leading-log contribution of $\lambda_1$ to the decay amplitude
 is
 \begin{equation} \label{hgg}
 {\it M}(h\to\gamma \gamma) = g_V^{(\prime)2}\frac{v_h \lambda_1\cos\theta_W^2}{8 \pi^2}
 g(x,z)\log \frac{\Lambda^2}{m_{V'}^2},\end{equation}
 where for the $g_V$ contribution $ g(x,z)\equiv
(13z-{5z^2}/{8}+{z}/{x}-3x^2)$  while for the $g_V^\prime$
contribution,   $ g(x,z)\equiv (13z-{5z^2}/{16}+{z}/{2x}-6x^2)$.
In both cases, $z=(m_H/m_V)^2$ and $x=(m_{V'}/m_V)^2$. Replacing
$\lambda_1 \rightarrow \lambda_2$ and $m_V \leftrightarrow
m_{V'}$, we obtain the contribution of $\lambda_2$. These
amplitudes have to be summed up with the SM triangle diagrams
within which top quark and $W$ boson propagate.
 Notice that our result is ultra-violet divergent. This is because
 $g_V$ and $\lambda_1$, despite being dimensionless, are
 non-renormalizable \cite{Weinberg}. For  $\lambda_1 \log \Lambda^2/m_{V'}^2\sim 1$, the
 contribution of $\lambda_1$ will be comparable to that in the SM.
 The observed slight excess \cite{LHC} can be attributed to this
 effect. Notice that if the excess is confirmed, the sign of $\lambda_1$ can also be determined.
 If further data rules out the excess, a bound on
 $\lambda_1 \log \Lambda^2/m_{V'}^2$ can be derived which for the
 value of $\lambda_1$ found in previous section
 ($\lambda_1={0.09}$) can be interpreted as an upper bound on
 $\Lambda$.  That is the scale of new physics giving rise to the effective $g_V$ coupling can be
 constrained. In case that $\lambda_3$ is the main contributor to
 the dark matter annihilation, the effect of $\lambda_1$ can be
 arbitrarily small. As discussed, these two possibilities can be
 distinguished by direct dark matter searches.

 With
 similar diagrams we predict a contribution to $H\to Z \gamma$.
 Since in this model the  new particles are heavier than $m_h$, the
 Higgs cannot have invisible decay modes.

 \section{Direct production at the colliders \label{LHC}}
 A pair of $V$ and $V^\prime$ can be produced by the annihilation of a fermion
 ($f$)
 and antifermion ($\bar{f}$) pair via a $s$-channel photon exchange,
 \small{\begin{eqnarray}
 \sigma (f \bar{f} \to V V^\prime)= \frac{g_V^{(\prime)2}(eQ_f \cos\theta_W)^2}
 {12\pi
 N_c E_{cm}^6m_V^2m_{V^\prime}^2}{\mathcal{K}}
 \mathcal{S}(E_{cm},m_V,m_{V^\prime})\label{ffbar}
 \end{eqnarray}
where
${\mathcal{K}}=\sqrt{(E_{cm}^2+m_V^2-m_{V^\prime}^2)^2-4m_V^2E_{cm}^2}$.
  For the $g_V$ contribution, we obtain
$$\mathcal{S}(E_{cm},m_V,m_{V^\prime})=[E_{cm}^2+2(m_V^2+m_{V^\prime}^2)][(E_{cm}-m_{V^\prime})^2-m_V^2][(E_{cm}+m_{V^\prime})^2-m_V^2]$$
while for the $g_V^\prime$ coupling $$
\mathcal{S}(E_{cm},m_V,m_{V^\prime})=[E_{cm}^4+(m_V^2-m_{V\prime}^2)^2](m_V^2+m_{V\prime}^2)-2E_{cm}^2(m_V^4-4m_V^2m_{V\prime}^2+m_{V^\prime}^4).$$
 Notice that the behavior of the cross section for $E_{cm}\to
\infty$ violates unitarity. This is because $g_V$ and
$g_V^\prime$} are  effective couplings below $\Lambda$. There is
also a subdominant contribution from $gg\to h^* \to VV^\prime$
which can be neglected relative to $ f\bar{f}\to \gamma^* \to
VV^\prime$. The energy of center in the LEP experiment was too low
to allow the production of $V$ and $V^\prime$ pair. However, in
the LHC, the $V$ and $V^\prime$ pair can be produced as long as we
are in the perturbative regime; {\it i.e.}, as long as
$m_{V^\prime}<$few TeV.

 Regardless of the mass range, $V^\prime$ can  decay to a photon and
 $V$. For $g_V^\prime=0$ and nonzero $g_V$
 $$\Gamma(V^\prime \to V
 +\gamma)= \frac{g_V^2}{96}\frac{
 \cos^2\theta_W}{\pi}\frac{(m^{2}_{V^\prime}-m_V^2)^3(m^{2}_{V^\prime}+m_V^2)}{m_V^2m^{5}_{V^\prime}}.$$
For $g_V=0$ and nonzero $g_{V}^\prime$, $g_V^2/96$ has to be
replaced with $g_V^{\prime 2}/24$. For both $g_V$ and $g_V^\prime$
regimes, the signature of the $V+V^\prime$ production at the LHC
will therefore be an energetic mono-photon plus missing energy
which has only low background \cite{Malik} and therefore enjoys a
good discovery chance. There is also a decay mode to $V+Z$
suppressed by $\tan^2 \theta_W$. If the kinematics allows
$V^\prime$ can decay to $V+H$, $V+W^-+W^+$ and $V+2H$; however,
the decay into $V+\gamma$ will dominate.
Using the parton distribution functions in \cite{pdf}, we have
calculated $\sigma(pp\to VV^\prime)$ and have found that for
$\sqrt{s}=7$~TeV and $m_{V'}=200~{\rm GeV}$  and for the value of
$g_V=0.19$ that  induces the desired 130 line intensity,
$\sigma(p+p\to V+V^\prime)=50$~fb which seems to be already
excluded by the 7 TeV run of the LHC \cite{Malik}. Thus, for
$g_V^\prime=0$, $m_{V'}$ should be larger than 200~GeV. For
$m_{V'}<500$ GeV and $g_V^\prime=0.4 (m_{V'}/500~{\rm GeV})$, the
cross-section $\sigma(p+p\to V+V^\prime)$ is larger than 50 fb so
$m_{V'}$ should be larger than 500~GeV. However, to draw a
conclusive result a dedicated analysis with customized cuts is
necessary. Taking $\sqrt{s}=8 ~{\rm TeV} (14 ~{\rm TeV})$ and
$m_{V^\prime}=1.5$~TeV and therefore $g_V=1.35$, we have found
$\sigma( p+p \to V+V^\prime) =0.5~{\rm fb}
 (90~{\rm fb}) $. Similarly, for the case of $g_V^\prime$ contribution
 with  $g_V^\prime=1.16$ and $m_{V^\prime}=1.5$~TeV,
  we have found
$\sigma(p+p\to
 V+V^\prime)=2~{\rm fb}
 (60~{\rm fb})$. Thus, the LHC can probe almost the whole perturbative
 regime.
Pairs of $V+V$  can be produced via $gg\to h^* \to VV$ at the LHC.
For $\lambda_1=0.09$, we have found the cross section to be {0.25}
fb ({0.8} fb) for 8 TeV (14 TeV) c.o.m energy.
\section{Conclusions \label{con}}
We have presented a model within which dark matter is composed of
a new vector boson ($V$) of mass 130~GeV such that through its
annihilation the observed 130 GeV photon line from the galaxy
center can be explained. The model also contains another vector
boson ($V^\prime$) which together they can couple to the
antisymmetric field strengths of   the photon and $Z$ boson. As
shown in Eq. (\ref{gvs}), two types of couplings are possible.
Both these couplings lead to the annihilation of dark matter pair
to two monochromatic lines: one line at 130 GeV and the other with
an intensity suppressed relative to the first one by $\sigma(VV\to
\gamma Z)/[2\sigma(VV\to \gamma \gamma)]<0.3$ at 114~GeV. Thus, by
searching for such double line feature the model can be tested.
$V^\prime$ has to have a mass smaller than a few TeV to account
for the 130 GeV line in the perturbative regime. The same coupling
can also lead to $V+V^\prime$ pair production at the LHC which
will appear as mono-photon plus missing energy signal. For a given
$V^\prime$ mass, the production rate is fixed. The present data
seems to already rule out light $V^\prime$. The entire
perturbative region with $m_{V'}<1.5$~TeV can be probed by the
14~TeV run of the LHC so this model is testable with this method,
too.

Within this model the dark matter pair mainly annihilates to a Higgs
pair with a cross section equal to 1~pb to account for the observed
dark matter abundance within the thermal dark matter scenario. This
annihilation can take place with either $\lambda_1$ coupling or the
$\lambda_3$ coupling defined in Eq.~(\ref{HV}). If $\lambda_1$ is
responsible for this annihilation, we expect an observable effect in
near future in direct searches for dark matter. In fact, the present
bound on dark matter-nucleon scattering cross section can be
accommodated only with small form factor. $\lambda_1$ can also
explain the small excess observed in $h\to \gamma \gamma$. It can
also contribute to  $h\to Z \gamma$. These observations can  fix the
sign of $\lambda_1$. However, if $\lambda_1\ll \lambda_3$, such
effects in the Higgs decay as well as direct dark matter searches
disappear.

The couplings that lead to dark matter pair annihilation to the
Higgs pair and $\gamma\gamma$ pair are all dimensionless.
Nonetheless, if the vector bosons are not gauge bosons, they will
be non-renormalizable leading to ultraviolet infinities and
violation of unitarity. Thus, these couplings are only effective
at low energies. However, as shown in \cite{non-decoupling}, the
``generalized Chern-Simons coupling'', $g_V^\prime$ can be large.
Using the St\"uckelberg mechanism, these vector bosons can be made
$U(1)$ gauge bosons, removing the cut-off dependence of $h\to
\gamma \gamma$ and violation of unitarity in the $V+V^\prime$
production at large center of mass energies.

If further data confirms the existence of the $\gamma$ line at 130
GeV, our model can provide a testable explanation with rich
phenomenology. If however this line disappears with further data
still the model has interesting features worth exploration.
Absence of any line would set an upper bound on $g_V$ and
$g_V^\prime$. If a photon line at a different energy appears, our
model with $m_V$ equal to the energy of the new line can provide
an explanation.
\section*{Acknowledgment}
The authors thank A. Smirnov and M. M. Sheikh-Jabbari for fruitful
discussion. They also thank R. Laha for useful comments. The
authors also thank the anonymous referee for her/his useful
comments. Y.F. acknowledges partial support from the European
Union FP7 ITN INVISIBLES (Marie Curie Actions, PITN- GA-2011-
289442).





\end{document}